\documentclass{emulateapj}

\begin{document}


\shorttitle{VUV measurements of interstellar CH}
\shortauthors{Sheffer \& Federman}

\title{{\it Hubble Space Telescope} Measurements of Vacuum Ultraviolet Lines of Interstellar CH}
\author{Y. Sheffer and S. R. Federman}
\affil{Department of Physics and Astronomy, University of Toledo, Toledo, OH 43606}
\email{ysheffe@utnet.utoledo.edu; steven.federman@utoledo.edu}

\begin{abstract}
Three interstellar absorption lines near 1370~\AA\ seen toward $\zeta$~Oph have been
assigned by Watson to Rydberg transitions in the $G$$-$$X$ (or $3d$$-$$X$) band of CH.
Our survey of a dozen diffuse interstellar lines of sight shows that the three absorption lines are
consistent with the known column densities of CH, by deriving the following
oscillator strengths:
$f$(1368.74) = 0.019 $\pm$ 0.003,
$f$(1369.13) = 0.030 $\pm$ 0.005, and
$f$(1370.87) = 0.009 $\pm$ 0.001.
We also determined intrinsic line widths that correspond to decay rates of
(1.5 $\pm$ 0.6) $\times$ 10$^{11}$ s$^{-1}$,
(3.8 $\pm$ 0.7) $\times$ 10$^{11}$ s$^{-1}$, and
(1.1 $\pm$ 0.6) $\times$ 10$^{10}$ s$^{-1}$
for $\lambda$$\lambda$1368, 1369, and 1370, respectively.
These rates are significantly higher than those associated with radiative decays, and thus are
readily attributable to predissociation of the Rydberg state.
A fourth interstellar line near 1271~\AA\ has been conjectured by Watson to be the strongest
transition in the $4d$$-$$X$ Rydberg band of CH.
We detected this line along four sight lines and our spectrum syntheses show that with
$f$(1271.02) = 0.007 $\pm$ 0.002, it is also consistent with the known column densities of CH.
In addition, we conducted a search for the $F$$-$$X$ band of CH near 1549~\AA, and
successfully discovered two of its absorption features along four sight lines.
The astronomical oscillator strengths derived for these features are
$f$(1549.05) = 0.021 $\pm$ 0.006 and
$f$(1549.62) = 0.013 $\pm$ 0.003.
Finally, the X~Per sight line provided us with an astronomical detection of another CH band,
via two $D$$-$$X$ features near 1694~\AA.
Comparisons with results of available theoretical calculations for the four CH bands are presented.
\end{abstract}


\section{Introduction}
The first listing of three unidentified (UID) features around 1369~\AA\ and along the `classic'
diffuse interstellar sight line toward $\zeta$~Oph was provided by \citet*{cse91}.
Their listing included two lines at 1368.74 and 1370.87~\AA, as well as the feature
$\lambda$1369.13, which according to their remark, resembled a diffuse interstellar band (DIB)
by virtue of its uncommon larger and asymmetric width.
This proposition was studied further by \citet*{tcs94}, who introduced the UID1 through UID3
nomenclature for these features and suggested that UID2 at 1369.13~\AA\ is probably the first
ultraviolet (UV) analog of the optical DIBs.
\citet*{gsk97} showed that as far as equivalent width
($W_{\lambda}$) is concerned, UID2 has a better correlation with $W_{\lambda}$(CO) than with
$W_{\lambda}$(\ion{Ni}{2}), or with \ion{H}{1} and \ion{Na}{1} column densities ($N$) along six
diffuse sight lines. 
Their conclusion was that UID2 most probably arises in a denser, molecular medium. 

\citet{watson01} performed a rigorous re-analysis of room-temperature laboratory data of CH
that was published by \citet{hj69}.
He showed that an assignment of the UID trio observed toward $\zeta$~Oph to CH is very plausible
on the grounds of wavelength
coincidence and approximate intensity or transition strengths.
Specifically, the assignment was to the $G$$-$$X$ band of CH, which was first observed in the
laboratory by \citeauthor{hj69}, who classified it as the the first member ($n$ = 3)
of the $nd$ Rydberg series.
Moreover, the defining characteristic of UID2, it being appreciably wider than other UID
and identified absorption lines, could be explained as arising from heterogeneous predissociation
of the Rydberg states of CH, including the $G$($^2\Sigma^+$, $^2\Pi$, $^2\Delta$), or 3$d$,
complex of three states.
The phenomenon of prominent Lorentzian wings signaling the extremely short lifetime of the
upper state is well known in Rydberg transitions of, e.g., CO \citep{viala88,eidelsberg06},
when closely-spaced rotational lines blend into each other more strongly in bands that
arise from states with faster predissociation rates, or wider wings.

Here we show that an identification of the UID features around 1369~\AA\ with $3d$$-$$X$
Rydberg transitions in CH is consistent with the expected range of $f$-values and with known
$N$(CH) values, as well as with line broadening arising from predissociation.
This is accomplished by utilizing profile synthesis to model the observed line shapes and
absorption strengths in high-resolution (high-$R$) {\it Hubble Space Telescope} ({\it HST}) spectra.
Furthermore, we were able to detect and analyze the $\lambda$1271 UID feature introduced by
\citet{federman95}, which \citet{watson01} conjectured to be the strongest transition in the
$4d$$-$$X$ band, the second member in the $nd$ Rydberg series of CH
\citep{hj69}.
Indeed, our spectral syntheses show that this UID is also consistent with theoretical
predictions for the band $f$-value, and with observed values of $N$(CH).

Finally, we conducted a deliberate search for additional spectral signatures of CH based on
the previous laboratory work of \citet{hj69}.
According to their Table 15, the bands $D$$-$$X$ at 1694~\AA, $E$$-$$X$ at 1557~\AA, and $F$$-$$X$
at 1549~\AA\ may be detected in the interstellar medium via their vacuum UV (VUV) transitions from
a $\Lambda$ component of the $J^\prime$$^\prime$ = 1/2 level of the $X~^2\Pi$ ground state.
Our search resulted in positive detections of the $F$$-$$X$ and the $D$$-$$X$ bands, but none of
the $E$$-$$X$ band.

A note concerning spectroscopic notation and molecular predissociation is in order.
The diatomic molecule's rotation is designated by $J$, with $N$ being its component that is
perpendicular to the internuclear axis of symmetry.
Radiative dipole transitions between lower ($J^\prime$$^\prime$,$N^\prime$$^\prime$) and upper
($J^\prime$,$N^\prime$) levels obey $\Delta$$J$ = $J^\prime$ $-$ $J^\prime$$^\prime$ = 0, $\pm$1.
Such transitions are designated by a primary letter $R$, $Q$, or $P$ for $\Delta$$J$ = +1,
0, or $-$1, respectively.
In addition, a superscripted $S$, $R$, $Q$, $P$, or $O$ stand for $\Delta$$N$ = $N^\prime$ $-$
$N^\prime$$^\prime$ = +2, +1, 0, $-$1, or $-$2, respectively.
For $\Delta$$N$ = $\Delta$$J$ transitions the superscripted letter will repeat
the primary letter, and is thus customarily omitted. 
Electronic states with angular momentum greater than 0, i.e., non-$\Sigma$ states, interact
with the nuclear spin to produce $\Lambda$-doubling of their rotational levels.
Thus the $X~^2\Pi$ ground state has two $\Lambda$ levels 0.11 cm$^{-1}$ (1.4 km s$^{-1}$)
apart, which are selectively connected by allowed transitions to upper states.
Figure 1 indicates the transitions analyzed here.
Owing to space limitations, $\Lambda$ levels and individual $J^\prime$ = $N^\prime$ $\pm$ 1/2
levels are not shown, while $N^\prime$ level separations are shown in a highly magnified and
schematic way.
Since we show those levels involved with transitions discussed here, only the lowest
$N^\prime$ level for each state is properly positioned with respect to the excitation energy axis. 
Once a transition to an upper electronic state has occurred, the strength of overlap between state
wavefunctions determines the probability of radiationless predissociation.
If the predissociation occurs via a state of similar electronic character, it is called homogeneous,
and the rate will not vary strongly with $J^\prime$ of the predissociated state.
Otherwise, heterogeneous predissociation occurs between states of differing characters, and
a strong dependence on $J^\prime$ is observed.

\section{Data and Its Modeling}
Our sample of 13 sight lines is obtained from spectra acquired with the Space Telescope Imaging
Spectrograph (STIS) and the Goddard High-Resolution Spectrograph (GHRS).
We analyzed the data, which are available from the {\it HST} archive, in order to derive $3d$$-$$X$
$f$-values for $\lambda$$\lambda$1368, 1369, and 1370 along 12 lines of sight.
We obtained smaller samples of sight lines to be analyzed in the search for additional
CH absorption features from its $4d$$-$$X$, $F$$-$$X$, and $D$$-$$X$ bands.
Whenever multiple exposures were available, we combined them in wavelength space for an
improved signal to noise (S/N) ratio.
In addition, when a feature (band) appeared in two adjacent orders, we combined them after
correcting for any small wavelength inconsistencies.
All our reductions were in IRAF and STSDAS.

Although it is known that the STIS line spread function (LSF) possesses weak wings, we could not
see any obvious signatures for such wings in the highest-$R$ data.
(The fact that $3d$$-$$X$ $\lambda$1369 looks wider than other lines shows that this characteristic
cannot be attributed to a global effect caused by prominent instrumental wings.)
Therefore, we used a single Gaussian to describe the instrumental profile of STIS,
in agreement with the findings of \citet{jt01}.
Nevertheless, the STIS handbook does show that the LSF wings have a larger role in the
profile for wider slits.
As a result, we did allow for $R$ to be a fitted parameter during modeling of
CO absorption lines (bands) taken from the same spectra used here.
We found that $R$ is lower for data taken through wider slits, and vice versa.
There is a good match between the range of resolving powers for a given slit and
the range found by \citet{bowers98}.
We will discuss these findings in a forthcoming publication about the CO molecule. 

Table 1 provides a list of sight lines in terms of the star observed,
its spectral type and Galactic direction, as well as the literature values for total $W_{4300}$(CH),
total $N$(CH), and cloud structure fractions along each line of sight.
For the first three sight lines, as well as the ninth, we use our latest $N$(CH) results based
on modeling with Ismod.f (see below) of high-$R$ ($\approx$ 170,000) optical spectra from McDonald
Observatory.
For the two stars in common, $o$ and $\zeta$ Per, our results agree within 5\% with
those of \citet*{cls95}.
A more detailed description of the new optical syntheses will be provided in a future publication.
All $N$(CH) determinations to date have utilized the $A~^2\Delta$$-$$X~^2\Pi$ 4300.313~\AA ``line'',
which is a $\Lambda$-doublet of two $R$(1/2) transitions only 1.4 km s$^{-1}$ apart \citep{bvd88}.
Table 2 lists the instrumental source for the VUV data, dataset names, exposure times,
S/N ratios, fitted values for the spectral resolving power, $R$ (from fits of CO absorption
present in the same spectra), and CH bands detected in the {\it HST} data.

All line profiles analyzed in this paper were modeled by spectrum synthesis
using Y. S.'s code, Ismod.f.
The basic radiative transfer equations in this code are taken from \citet{bvd88}.
Fit parameters include radial velocity, $N$(CH), $f$-value, line width, and cloud component
structure, which involves relative velocities, fractional abundances, and Doppler $b$-values.
Both $N$(CH) and the cloud component structure were known a priori for each sight line
and were kept fixed during the CH fits.
Ismod.f employs the Voigt profile to describe the line shape as a combination of a Gaussian
Doppler core and Lorentzian wings.
The resulting profile is then convolved with the appropriate instrumental Gaussian
profile before a comparison with the data is performed via root-mean-square computation.
Ismod.f employs the simplex method to converge onto the final fit, which is found by an automated
trial and error inspection of the parameter space in decreasing steps, until a certain
tolerance level (usually 10$^{-4}$ of the parameter's value) has been reached.
Figure 2 shows the data and Ismod.f profile syntheses of the three $3d$$-$$X$ lines toward the star
X~Per.

\section{Individual Sight Lines}
We first compare a pertinent selection of our results with previously published values.
Most of the work done previously concerned the well-known sight line toward $\zeta$~Oph.
Our results from profile syntheses of $\lambda$$\lambda$1368, 1369, and 1370 show that fitted
$W_{\lambda}$ values for the three lines are 5.6 $\pm$ 0.2, 7.5 $\pm$ 0.2, and 2.2 $\pm$ 0.3 m\AA,
respectively, where our uncertainties reflect noise contribution over the extent of the profile
from the expression FWHM/(S/N).
When compared to the two previous publications, our $W_{1368}$ and $W_{1370}$ agree more with
\citet[4.02 $\pm$ 0.58, 4.59 $\pm$ 0.74, and 1.16 $\pm$ 0.39 m\AA, respectively]{cse91},
while our $W_{1369}$ agrees more with
\citet[3.80 $\pm$ 0.80, 5.89 $\pm$ 0.89, and 0.84 $\pm$ 0.59 m\AA, respectively]{tcs94}.
It is possible that our $W_{\lambda}$ values are larger because they include more of the
contribution of the wide wings that arise in predissociation.
When wide wings are unknowingly present, their depression of the local continuum might be
assumed to arise in the continuum itself (e.g., owing to stellar undulations).
Special care should be taken to preserve such wings during the rectification process,
but this is hard to accomplish for data with low S/N.

\citet{gsk97} measured $W_{1369}$ toward six stars, of which three
sight lines are in common with this study.
Therefore, we may compare our synthesized $W_{\lambda}$ values (in m\AA) with those of
\citeauthor{gsk97}:
X~Per, 8.0 $\pm$ 0.3 vs. 8.4; $\zeta$~Oph, 7.5 $\pm$ 0.2 vs. 6.3; and $\lambda$~Cep,
7.5 $\pm$ 0.5 vs. 6.9. 
There are no major disagreements between the two sets of values.
Note that for the $\zeta$~Oph sight line, the value from \citeauthor{gsk97} is also
higher than those from \citet{cse91} and \citet{tcs94}.

\citet{federman95} have been the only authors to report the presence of the
$\lambda$1271 feature, in the direction of $\zeta$~Oph.
Their reported value was $W_{\lambda}$ = 0.75 $\pm$ 0.19 m\AA, while our fitted result is
1.1 $\pm$ 0.1 m\AA. The latter agrees within 2 $\sigma$ of the earlier value.

\section{The $3d$$-$$X~^2\Pi$ band of CH}
\subsection{Oscillator Strengths}
An absorption line is produced by the combined action of $N$ absorbers
per cm$^2$ along the line of sight, each removing photons from the beam at the
transition wavelength ($\lambda$ in \AA) at a rate that corresponds to an oscillator strength,
or $f$-value.
The mutual product $fN$ appears in the equation that determines the optical depth
in the transition, which then generates the $W_{\lambda}$ of the line profile through
exponential absorption along the line of sight.
It is customary to have at hand transitions with known $f$-values that consequently allow
the determination of $N$ of the absorbers.
Here, on the other hand, we use the known values of $N$(CH) from 12 sight lines to determine the
$f$-values for $\lambda$$\lambda$1368, 1369, and 1370.

Table 3 provides the results of our spectrum syntheses for the three $3d$$-$$X$ lines toward our
target stars: fitted $W_{\lambda}$, $f$-values, and inferred intrinsic Lorentzian line widths
(see more about the width in the next section).
Normally, $N$(CH) is determined from the $A$$-$$X$ doublet of $J^\prime$$^\prime$ = 1/2 transitions
at 4300.313~\AA, where the two blended $\Lambda$ levels have their populations combined.
Indeed, ultra-high-$R$ spectra at 4300~\AA\ \citep*[e.g.,][]{lsc90}
sometimes resolve the 1.4 km s$^{-1}$ doublet, revealing equal populations for the two components.
Here, we emphasize that each of our profile syntheses employs only half the value of $N$(CH)
(taken from Table 1) because in all resonance transitions to higher (than $A$) electronic states
of CH, the individual $\Lambda$ components are well separated \citep[see][]{lien84}. 
Therefore, the $f$-values derived here are twice as large as they would have been, had the
full value of $N$(CH) been used.
Within the uncertainties listed, there is good agreement among all sight lines
for each $f$-value.
Non-weighted means for the three lines are $f$(1368) = 0.019 $\pm$ 0.003,
$f$(1369) = 0.030 $\pm$ 0.005, and $f$(1370) = 0.009 $\pm$ 0.001.
For these averages we used fit results from the higher-$R$ STIS spectra,
while avoiding 2 (in the case of $\lambda$1369) or 3 (otherwise) sight lines from lower-$R$ GHRS
exposures.
Low-$R$ data do not provide as good a handle on intrinsic line profiles as do
high-$R$ data, both in the reduction (i.e., continuum rectification) stage and in the
modeling stage.
Had we used the excluded results, the average $f$-values would have
increased by 11\%, 10\%, and 3\%, respectively.
Thus the sum of the STIS-derived $f$-values for the three transitions is 0.058 $\pm$ 0.006, which
according to \citet{watson01} is providing some 76\%  of the total $f$-value of the
$3d$$-$$X$ band, because the three lines are the most prominent transitions in it.
With the correction, this Rydberg band has a total $f$-value of 0.076 $\pm$ 0.008, as
inferred by our interstellar measurements.

\citet{watson01} derived a preliminary $f$-value for the $3d$$-$$X$ band from the sum of 
$W_{\lambda}$ of the three lines in \citet[10.5 m\AA]{tcs94}, employing $N$(CH) = 10$^{13.4}$
cm$^{-2}$ from \citet{fl88}, and the expression for unsaturated line absorption,
$log (W_{\lambda}/\lambda) = log (f\lambda) + log (N) - 20.053$, from \citet{mn94}.
From this value, $f$ = 0.025, which was not based on any profile spectral synthesis,
\citet{watson01} estimated the total band $f$-value to be 0.033.
(We estimate that the uncertainties associated with \citeauthor{watson01}'s $f$-values amount to
20\%, based on the $W_{\lambda}$ uncertainties in \citeauthor{tcs94})
Both values appear to be too low by a factor of 2.3 $\pm$ 0.5 when compared to our results.
However, Watson's determination was based on the full value of the $N$(CH),
whereas in our modeling we use $N$(CH)/2, which is pertinent for the population
of a single $\Lambda$-doubling level, see above.
Correcting \citeauthor{watson01}'s estimate upward by a factor of 2, we get $f$ $\approx$
0.050 $\pm$ 0.010 for the sum of three lines, and $f$ $\approx$ 0.066 $\pm$ 0.013 for the entire
$3d$$-$$X$ band.
This happens to be in much better agreement with our rigorous spectrum synthesis results.
Only two theoretical ab initio band $f$-values exist in the literature: 0.0467 from
\citet[not 0.0362 as per \citeauthor{watson01}]{bn78}, and 0.069 from \citet{vandishoeck87}.
It appears that both our astronomically-derived band $f$-value, and the \citet{watson01} value
when corrected upward by a factor of 2, are in good agreement with the theoretical value
calculated by \citet{vandishoeck87}, thus bolstering the case for identifying these three
lines with the CH molecule.
This measurement also shows $3d$$-$$X$ to be the CH band with the largest $f$-value.
Perhaps this is what \citet{hj69} meant when they commented that $G$$-$$X$ lines
would be the strongest interstellar lines of CH.

\subsection{Predissociation Widths}
As mentioned in the previous section, Table 3 provides information about the Lorentzian line
width derived in our modeling.
The intrinsic line width is set by the lifetime of the transition's upper level, according to the
inverse quantum mechanical relationship between lifetime (limiting the interval of measurement)
and line width (resulting uncertainty in measured energy).
Normally, intrinsic line widths are unobservable owing to the relatively ``long'' lifetimes
of radiative transitions; $\tau \geq$ 10$^{-8}$ s yields a width of 0.01 m\AA\ or less, using
the relationship
$ \Gamma(\textrm{m\AA}) = [\lambda(\textrm{\AA})]^2/(1.884\times10^{16}\tau). $
Predissociation generates much shorter lifetimes for certain Rydberg states,
owing to decay rates that may readily be 10$^3$ times faster than radiative decay rates.
Under such circumstances, the associated lifetime would be 10$^{-11}$ s, leading to line widths
of the order of 10 m\AA,
which can be measured on high-$R$ ($\approx$ 100,000) spectra with the help of profile synthesis. 
We estimate that in our highest-$R$ data it is possible to reliably extract
$\Gamma$ widths of a few m\AA, which correspond to decay rates that are higher than a few $\times$
10$^{10}$ s$^{-1}$.
For $f$-value syntheses that failed to converge onto a reliable $\Gamma$ solution, we fixed the
line width to a value near the average from well-converged solutions.
The predissociation decay rate, $k_{pr}$, is the inverse of $\tau$, and thus
is directly proportional to $\Gamma$.
We use both $k_{pr}$ and $\Gamma$ as descriptors of line width in terms of Lorentzian wings.

Modeling the three $3d$$-$$X$ lines with Ismod.f returned values for the Lorentzian width in
inverse lifetimes (s$^{-1}$), which are tabulated in Table 3.
We computed the means from all sight lines with high-$R$ (non-GHRS)
determinations and derived the following predissociation rates:
$k_{pr}$(1368) = (1.5 $\pm$ 0.3) $\times$ 10$^{11}$ s$^{-1}$,
$k_{pr}$(1369) = (3.8 $\pm$ 0.7) $\times$ 10$^{11}$ s$^{-1}$, and
$k_{pr}$(1370) = (1.1 $\pm$ 0.6) $\times$ 10$^{10}$ s$^{-1}$.
It can be seen that all data sets returned a well-defined line width for $\lambda$1369,
thanks to its larger $\Gamma$, but that only a few syntheses were successful in converging
onto a $\Gamma$ solution for $\lambda$1370, the profile with the narrowest wings.
In fact, whereas average $k_{pr}$ values for both $\lambda$1368 and $\lambda$1369 are better than the
2.5 $\sigma$ detection level, $k_{pr}$ for $\lambda$1370 is a $<$ 2 $\sigma$ result.
It is thus possible that the width of $\lambda$1370 has not been reliably determined, and that a
3 $\sigma$ upper limit of $\approx$ 3 $\times$ 10$^{10}$ s$^{-1}$ for $k_{pr}$(1370) should be used
instead.

According to the $\Gamma$--$\tau$ relationship, the corresponding Lorentzian profile widths
of the three $3d$$-$$X$ lines
are 11.6, 38.4, and 1.1 m\AA, which correspond to 0.62, 2.05, and 0.06 cm$^{-1}$.
\citet{watson01} used the value 2.6 cm$^{-1}$ for the width of $\lambda$1369, which was based on the
value of 10.7 km s$^{-1}$ in \citet{tcs94}.
We also note that the original observation of $\lambda$1369 toward $\zeta$~Oph was described as
showing an asymmetric profile.
Here, our impression from the entire sample of sight lines is that the observed profiles of
$\lambda$1369 are very consistent with formation by a single transition, i.e., there are no
indications for asymmetries in the line profile, especially in higher-$R$ data.

Theory has no definite bearing on these $k_{pr}$ values, yet. \Citet{vandishoeck87} remarked that
although predissociation is present in all bands studied by \citet{hj69}, for most
bands, including the $3d$$-$$X$ band at 1370~\AA, it is heterogeneous.
Thus the predissociation rate may approach an estimated value of 10$^{10}$ s$^{-1}$ for
$N^\prime$ = 10, but is estimated to be about 10$^{8}$ s$^{-1}$ for low $N^\prime$, i.e., of the
order of the radiative decay rate of the state.
These estimates are not rigorous enough to be more accurate that an order of magnitude. 
Indeed, our fits and \citeauthor{watson01}'s initial probe do show that in the case of both
$\lambda$1368 and $\lambda$1369 there is a substantial contribution to CH photodissociation, since
their $k_{pr}$ values are much higher than what \citet{vandishoeck87} had suggested.
According to \citet{watson01}, all three lines are $N^\prime$ = 2 transitions.

Figure 2 shows that there is good agreement between our models and observed line profiles.
For $\lambda$1369, there is no need to invoke an analogy to the DIBs in the optical, because we see
that very good fits are obtained using a single transition with broad predissociation Lorentzian
wings.
In addition to agreeing with \citet{watson01} that CH is a very good assignment for
$\lambda$$\lambda$1368, 1369, and 1370, one may also conclude that VUV analogs of the optical DIBs
have yet to be detected.

\section{The $4d$$-$$X~^2\Pi$ band of CH}
Our inspection of archival STIS data returned four candidate sight lines with $\lambda$1271
features that appear to be astronomical detections of the strongest transition in $4d$$-$$X$.
The sight lines are toward
X~Per (shown in Figure 3), $\zeta$~Oph, HD~203532, and HD~207308, as listed in Table 4.
Despite varying values of S/N and $R$, all four spectrum syntheses returned rather consistent
solutions, $f$ = 0.007 $\pm$ 0.002.
This $f$-value is smaller than that for $\lambda$1369, the strongest $3d$$-$$X$ transition with
$f$ of 0.030 $\pm$ 0.005 reported above.
A theoretical prediction from \citet{bn78} gives $f$-values of 0.0198 and 0.0186
for the $^2\Delta$ and $^2\Pi$ states in the $4d$ complex, respectively.
If we assume that the total band $f$-value is 2.5 times the value of its strongest transition,
as we find for the $3d$$-$$X$ band complex and $\lambda$1369, then from the $f$-value for
$\lambda$1271 we expect $f$ $\approx$ 0.0175 for the entire $4d$$-$$X$ band.
This value is in close agreement with the predictions of \citeauthor{bn78} for
a single ($^2\Delta$ or $^2\Pi$) state only, falling short by a factor of 2 relative to their sum.
\Citet{vandishoeck87} did not include the $4d$ state in her $f$-value calculations.

However, \citeauthor{vandishoeck87} suggested that $k_{pr}$($4d$) would be smaller than that of
the $3d$ state owing to a smaller overlap of the $4d$ state with the nuclear wavefunctions.
Three sight lines with either the highest $R$ or the highest S/N values yielded solutions for
$\Gamma$, while for HD~207308 the fit did not converge onto a reliable Lorentzian line width.
With an average of $k_{pr}$ = (2.5 $\pm$ 1.5) $\times$ 10$^{11}$ s$^{-1}$, this transition has
a predissociation rate between those of $\lambda$1368 and $\lambda$1369.
Owing to small number (n = 3) statistics, this $<$ 2 $\sigma$ result seems to suffer
from either noise and/or continuum biases.
Furthermore, because only a single $J^\prime$$^\prime$ = 1/2 transition has been found so far,
the detection of $4d$$-$$X$ is not as secure as those of the other CH bands described here.

\section{The $F~^2\Sigma^+$$-$$X~^2\Pi$ band of CH}
\citet{hj69} observed this band at 1549~\AA, i.e., it is located between the
two components of the \ion{C}{4} doublet $\lambda$$\lambda$1548, 1550 seen in stellar and
interstellar spectra.
They noticed an increasing diffuseness with higher $J$, signaling heterogeneous predissociation.
We detected two $J^\prime$$^\prime$ = 1/2 absorption features from the $F$$-$$X$ band
along the sight lines toward X~Per (see Figure 4), HD~147683, $\zeta$~Oph, HD~207308, and
$\lambda$~Cep, as listed in Table 4.
The redder, weaker line at 1549.62~\AA\ belongs to a single transition, $^PQ_{12}$, while the
stronger line at 1549.05~\AA\ is an unresolved blend of the two transitions $Q_2$ and $^QR_{12}$.
According to \citet{lien84}, who analyzed the $C~^2\Sigma^+$$-$$X~^2\Pi$ band of CH, both $Q_2$
and $^PQ_{12}$ have a H\"{o}nl-London factor (HLF) of 4/3, while $^QR_{12}$ has HLF = 2/3.
(Since the $F$ state is also of a $^2\Sigma^+$ character, the same HLFs should apply in
$F$$-$$X$ and $C$$-$$X$ transitions.)
The line of sight toward HD~23478 shows only the strongest feature from
the $F$$-$$X$ band, i.e., the blend $\lambda$1549.05.

The averages of our fitted $f$-values, based on the known values of $N$(CH)/2, are
$f$(1549.05) = 0.021 $\pm$ 0.006 and $f$(1549.62) = 0.013 $\pm$ 0.003.
The $f$-value ratio is 1.6 $\pm$ 0.6 for $\lambda$1549.05 vs. $\lambda$1549.62.
Based on the actual transitions that constitute these lines and their individual HLF
values, the predicted $f$-value ratio is 1.5, showing good agreement with our measurements.
The total $f$-value of $\lambda$1549.05 equals the total band $f$-value, according to
the sum of HLFs, or 4/3 + 2/3 = 6/3. 
Here we derive a sum of $f$ = 0.021 $\pm$ 0.006, whereas \citet{bn78} and
\citet{vandishoeck87} theoretically computed values of 0.031 and 0.0095, respectively.
Our value lies between the predicted $f$-values, about a factor of 2 from either one.

There is another potentially detectable $J^\prime$$^\prime$ = 1/2 absorption line
belonging to the $F$$-$$X$ band, $R_2$ at 1547.87~\AA\ \citep{hj69}.
We were not able to detect it owing to
blending with the blue wing of the much stronger line of \ion{C}{4} at 1548.204~\AA.
According to its HLF of 2/3, the $R_2$ transition would constitute the weakest
$J^\prime$$^\prime$ = 1/2 absorption line.
Since we measure two lines composed of the three strongest transitions with verified $f$-value
ratios, these first astronomical detections of the $F$$-$$X$ band of CH seem secure. 

We also extracted Lorentzian widths for the absorption lines of the $F$$-$$X$ band,
as listed in Table 4.
As is the case for the lines of the $3d$$-$$X$ band, under conditions of high-$R$
the lines readily show that different predissociation widths are involved, confirming
\citeauthor{hj69}'s observation of heterogeneous predissociation for $F$$-$$X$.
Also in a similar fashion to $3d$$-$$X$, the strongest line in the band appears to be the widest.
Thus, for the blend $Q_2$ + $^QR_{12}$ at 1549.05~\AA, we find $k_{pr}$ = (2.7 $\pm$ 1.3) $\times$
10$^{11}$ s$^{-1}$, whereas $k_{pr}$ for $^PQ_{12}$ at 1549.62~\AA\ is (1.6 $\pm$ 0.9) $\times$
10$^{10}$ s$^{-1}$.
A Golden Rule approximation by \citet{vandishoeck87} yielded a computed $k_{pr}$ of 7.6 $\times$
10$^{9}$ s$^{-1}$ for the $N^\prime$ = 1 levels of the $F$ state, with a stated uncertainty
of ``an order of magnitude''.
The two transitions with $N^\prime$ = 1 ($Q_2$ and $^QR_{12}$) are those that make up the blend
at $\lambda$1549.05, for which we find $k_{pr}$ that is larger by 1.5 orders of magnitude than
the approximate prediction.

We attempted a more detailed spectrum synthesis of $\lambda$1549.05 toward X~Per that
explicitly included the two transitions $Q_2$ and $^QR_{12}$, subject to the constraint of a 2:1
ratio in their HLF values.
Although the fit returned a separation of $-$6.2 km s$^{-1}$ for the weaker $^QR_{12}$ relative to
$Q_2$, we take this to be an upper limit because the fit yielded a solution with
residuals that are somewhat smaller that the ambient noise level away from the feature.
The actual energy level splitting in the $F~^2\Sigma^+$ state that controls the velocity
separation between the two $\lambda$1549.05 transitions is unknown, but we note that the $C$$-$$X$
band has its $^QR_{12}$ line located 2.2 km s$^{-1}$ to the blue of $Q_2$ \citep{lien84}.

This detailed synthesis of the $Q_2$ + $^QR_{12}$ blend resulted in narrower Lorentzian wings for
the two transitions and (predictably) lower $f$-values relative to the single-transition model.
It returned $f$($Q_2$) = 0.010 and $f$($^QR_{12}$) = 0.005, i.e., a sum that is
12\% smaller than the $f$-value fitted for the entire blend, and
decay rates $k_{pr}$ = 1.3 $\times$ 10$^{11}$ and 1.1 $\times$ 10$^{11}$ s$^{-1}$, for $Q_2$
and $^QR_{12}$, respectively.
These lower $k_{pr}$ values are still higher than the approximate prediction
of \citet{vandishoeck87}.
The similarity in decay rates for the two transitions is inconsistent with their
$J^\prime$ levels of 1/2 (for $Q_2$) and 3/2 (for $^QR_{12}$), because
heterogeneous predissociation is normally revealed by $\Gamma$ having a dependence on $J$($J$+1),
i.e., we expect $^QR_{12}$ to be 5 times wider than $Q_2$.
Since detailed modeling attempts for other sight lines did not return separations and widths
for the blended components that were in good agreement with the X~Per result,
the quality of the data must be insufficient for such detailed modeling,
especially when the line separation is not available as a known input value.

\section{The $D~^2\Pi$$-$$X~^2\Pi$ band of CH}
This band was observed by \citet{hj69}, who remarked on the relative weakness
of its $Q$-branch and the apparent diffuse nature of its lines.
A single sight line, that toward X~Per, exhibits an absorption feature that is consistent
with a detection of $\lambda$1695.34, the strongest feature in $D$$-$$X$ (see Figure 5 and Table 4).
Here the strongest feature is also an unresolved blend
of the two transitions $Q_2$ and $^QR_{12}$ \citep{kt73}.
Furthermore, a shallower feature in the spectrum is consistent with the position of the
$R_2$ $\lambda$1693.24 transition of $D$$-$$X$, although it is partially blended with a stronger
\ion{Si}{1} $\lambda$1693.29 line, see Figure 6.
Upon modeling, the known $N$(CH)/2 toward X~Per returns $f$-values of 0.0038 $\pm$ 0.0011 and
0.0012 $\pm$ 0.0006 for $Q_2$ + $^QR_{12}$ and $R_2$, respectively, while the ratio
of their $f$-values is 3.2 $\pm$ 1.8.
\citet{kt73} theoretically predicted a 5:1 ratio
for $\lambda$1695.34 vs. $\lambda$1693.24, a value that is consistent with our determinations.
Theoretical computations by \citet{vandishoeck87} gave $f$ = 0.0073 for the 0$-$0
band of the $^2\Pi_1$ state, which she identified with the $D$ state.
\citet{bn78} calculated a value of 0.0010 for $f$($D$$-$$X$).
Our value of 0.0038 $\pm$ 0.0011 for the strongest feature appears to be within a factor of
2 to 3 of the two quantum mechanical predictions.
In light of the good agreement in wavelength and $f$-value with predicted values,
we have high confidence in treating this case as the first astronomical
detection of the $D$$-$$X$ band from CH.

Concerning Lorentzian line width, our synthesis of the blend at $\lambda$1695.34 returned
$k_{pr}$ = 7.0 $\times$ 10$^{10}$ s$^{-1}$, 
a rate which was kept fixed during the modeling of the partially blended $R_2$ line.
However, \citet{ll99} measured decay rates of (3.6 $\pm$ 0.6) $\times$ 10$^{11}$ s$^{-1}$
in the laboratory for the $N^\prime$ = 2 levels of $D$$-$$X$, and \citet{mm00} calculated
a life time of 1.6 ps for the same state, albeit they denoted it by the letter $E$.
These consistent non-astronomical values show that predissociation dominates the decay
of $D$ levels, as happens for the $3d$ levels.
Consequently, we re-synthesized $\lambda$1695.34 with a fixed input $\Gamma$ value from
\citet{ll99}, yielding $f$ = 0.0057 $\pm$ 0.0023, which is in good agreement with the
calculated value of 0.0073 from \citep{vandishoeck87}.
Likewise, the $f$-value of $\lambda$1693.24 also increased by about 50\%, see Table 4.
According to Figure 5, it seems that owing to insufficient S/N our original fit of $\lambda$1695.34
had converged onto a solution involving a noise glitch, leading to a $k_{pr}$ value that was
too small.
The revised fits with the terrestrial value for $\Gamma$ are also consistent with the data.

\section{Conclusions}
We surveyed a sample of interstellar sight lines and specific spectral ranges to search
for new astronomical detections of VUV transitions of the CH molecule, as well as
to confirm \citet{watson01}'s identification of four UID lines as transitions involving
two Rydberg states of CH.
We found that both old and new detections appear to be consistent with the identification
of CH as their source, based on known amounts of CH along the lines of sight, on predicted
wavelength positions, and on theoretical $f$-values. 
Thus the decades-long tradition of optical observations of CH, utilizing its $A$$-$$X$,
$B$$-$$X$, and $C$$-$$X$ bands, can now be extended into the VUV via observations of its
$D$$-$$X$, $F$$-$$X$, $3d$$-$$X$ ($G$$-$$X$) bands, and, more reservedly, the $4d$$-$$X$ band.

We could not find any absorption features that might be identified with $E$$-$$X$ transitions of CH,
listed by \citet{hj69} to be near 1557~\AA.
This band is problematic with an unknown configuration, and could be arising either from an
$E~^2\Pi$ state or it could be the 2$-$0 band of $D$$-$$X$ \citep*{ccc85}, which has a very
low $f$-value of 3.0 $\times$ 10$^{-6}$ \citep{vandishoeck87} and a very high rate of
predissociation, $k_{pr}$ = 1.4 $\times$ 10$^{13}$ s$^{-1}$ \citep{mm00}.
Both characteristics of extremely weak oscillator strengths and extremely wide line profiles
would easily render this band undetectable in astronomical, or low-$N$(CH), settings.

Finally, with the confirmation that wider-than-average line profiles are attributable to
high predissociation rates and symmetric Lorentzian wings in Rydberg transitions of CH,
there are no candidates currently left in the VUV spectral regime that may be associated with
the famous optical DIB bands. 

\acknowledgments
We acknowledge support from NASA (grants NAG5-11440 and NNG06GC70G) and the Space Telescope
Science Institute (grant HST-AR-09921.01-A). We thank Dr. D. Welty for providing us with
CH cloud structures toward X~Per and HD~154368 prior to publication and an anonymous
referee for helpful comments.

\begin{deluxetable}{lllrrcccc}
\tabletypesize{\scriptsize}
\tablewidth{0pc}
\tablecaption{Stellar and CH Data for Program Stars\tablenotemark{a}}
\tablehead{
\colhead{Star}
&\colhead{Name}
&\colhead{Sp.}
&\colhead{$l$}
&\colhead{$b$}
&\colhead{$W_{4300}$(CH)}
&\colhead{$N_{\rm tot}$(CH)}
&\colhead{$N_{\rm i}$(CH)/$N_{\rm tot}$(CH)}
&\colhead{CH Ref.\tablenotemark{b}}\\
\colhead{}
&\colhead{}
&\colhead{}
&\colhead{($^\circ$)}
&\colhead{($^\circ$)}
&\colhead{(m\AA)}
&\colhead{(10$^{13}$ cm$^{-2}$)}
&\colhead{}
&\colhead{}}
\startdata
HD 23180  &$o$ Per     &B1III  &160.36 &$-$17.74 &14.5 $\pm$ 0.5 &1.90 $\pm$ 0.06 &.37, .63           &1 \\
HD 23478  &            &B3IV   &160.76 &$-$17.42 &14.1 $\pm$ 0.5 &1.83 $\pm$ 0.07 &.74, .26           &1 \\
HD 24398  &$\zeta$ Per &B1Iab  &162.29 &$-$16.69 &15.7 $\pm$ 0.1 &2.14 $\pm$ 0.01 &.56, .44           &1 \\
HD 24534  &X Per       &O9.5pe &163.08 &$-$17.14 &26.3 $\pm$ 0.9 &3.69 $\pm$ 0.13 &.14, .16, .70      &2 \\
HD 147683 &V760 Sco    &B4V    &344.86 &  +10.09 &18.4 $\pm$ 2.2 &2.22 $\pm$ 0.26 &.09, .68, .23      &3 \\
HD 149757 &$\zeta$ Oph &O9V    &  6.28 &  +23.59 &19.6 $\pm$ 2.0 &2.50 $\pm$ 0.25 &.50, .08, .42      &4 \\
HD 154368 &V1074 Sco   &O9Ia   &349.97 &   +3.22 &38.7 $\pm$ 1.3 &6.25 $\pm$ 0.21 &.03, .01, .03, .18, .70, .04, .01 &2 \\
HD 203374 &            &B0IVpe &100.51 &   +8.62 &17.3 $\pm$ 1.0 &2.22 $\pm$ 0.11 &.05, .04, .32, .20, .28, .06, .05 &5 \\
HD 203532 &            &B3IV   &309.46 &$-$31.74 &18.0 $\pm$ 0.6 &2.46 $\pm$ 0.08 &.40, .60           &1 \\
HD 207308 &            &B0.5V  &103.11 &   +6.82 &23.7 $\pm$ 1.0 &3.21 $\pm$ 0.16 &.07, .09, .57, .27 &5 \\
HD 207538 &            &O9V    &101.60 &   +4.67 &28.0 $\pm$ 1.2 &3.81 $\pm$ 0.19 &.06, .18, .31, .31, .11, .03 &5 \\
HD 208266 &            &B1V    &102.71 &   +4.98 &25.5 $\pm$ 1.2 &3.35 $\pm$ 0.17 &.01, .39, .37, .12, .11 &5 \\
HD 210839 &$\lambda$ Cep&O6Iab &103.83 &   +2.61 &16.5 $\pm$ 1.7 &2.08 $\pm$ 0.21 &.38, .62           &4 
\enddata
\tablenotetext{a}{We used the SIMBAD database for spectral types and Galactic coordinates.}
\tablenotetext{b}{
(1) This paper;
(2) Welty 2005, priv. comm.;
(3) \citealt*{awc02};
(4) \citealt{cls95};
(5) \citealt{pan04}.
}
\end{deluxetable}

\begin{deluxetable}{llllrrrl}
\tabletypesize{\scriptsize}
\tablewidth{0pc}
\tablecaption{{\it HST} Spectroscopy of Program Stars}
\tablehead{
\colhead{Star}
&\colhead{Instr.}
&\colhead{Dataset}
&\colhead{Grating}
&\colhead{Slit}
&\colhead{S/N}
&\colhead{$\lambda/\Delta\lambda$}
&\colhead{CH bands}\\
\colhead{}
&\colhead{}
&\colhead{}
&\colhead{}
&\colhead{(\arcsec)}
&\colhead{}
&\colhead{}
&\colhead{}}
\startdata
HD 23180  &STIS &o64801--4  &E140H &0.2X0.05 &100 &120000 &$3d$$-$$X$ \\
HD 23478  &STIS &o6lj01     &E140H &0.1X0.03 &30  &130000 &$3d$$-$$X$, $F$$-$$X$ \\
HD 24398  &STIS &o64810--11 &E140H &0.2X0.05 &90  &120000 &$3d$$-$$X$ \\
HD 24534  &STIS &o64812--13 &E140H &0.1X0.03 &105 &140000 &$3d$$-$$X$ \\
          &STIS &o66p01     &E140H &0.2X0.09 &85  &100000 &$4d$$-$$X$, $3d$$-$$X$ \\
          &STIS &o66p02     &E140H &0.2X0.09 &55  &100000 &$F$$-$$X$ \\
          &STIS &o66p03     &E140H &0.2X0.09 &110 &100000 &$D$$-$$X$ \\
HD 147683 &STIS &o6lj06     &E140H &0.2X0.09 &35  &120000 &$3d$$-$$X$, $F$$-$$X$ \\
HD 149757 &GHRS &z0x201     &G160M &0.25     &900 & 20000 &$4d$$-$$X$, $3d$$-$$X$ \\
          &GHRS &z0hu01     &G160M &2.0      &110 & 15000 &$F$$-$$X$ \\
HD 154368 &GHRS &z0wx02     &G160M &0.25     &30  & 20000 &$3d$$-$$X$ \\
HD 203374 &STIS &o6lz90     &E140M &0.2X0.2  &45  & 40000 &$3d$$-$$X$ \\
HD 203532 &STIS &o5c01s     &E140H &0.2X0.2  &25  & 80000 &$4d$$-$$X$ \\
HD 207308 &STIS &o63y02     &E140M &0.2X0.06 &60  & 45000 &$4d$$-$$X$, $3d$$-$$X$, $F$$-$$X$ \\
HD 207538 &STIS &o63y01     &E140M &0.2X0.06 &55  & 45000 &$3d$$-$$X$ \\
HD 208266 &STIS &o63y03     &E140M &0.2X0.06 &60  & 45000 &$3d$$-$$X$ \\
HD 210839 &GHRS &z12j01     &G160M &0.25     &230 & 20000 &$3d$$-$$X$ \\
          &GHRS &z2qx01--04 &G160M &2.0      &210 & 15000 &$F$$-$$X$
\enddata
\end{deluxetable}

\begin{deluxetable}{lccccccccccc}
\tabletypesize{\scriptsize}
\tablewidth{0pc}
\tablecaption{Astronomical Measurements of $3d$$-$$X$ lines of CH}
\tablehead{
\colhead{Star}
&\colhead{W$_{\lambda}$}
&\colhead{$f$-value}
&\colhead{$k_{pr}$/10$^{11}$}
&\colhead{}
&\colhead{W$_{\lambda}$}
&\colhead{$f$-value}
&\colhead{$k_{pr}$/10$^{11}$}
&\colhead{}
&\colhead{W$_{\lambda}$}
&\colhead{$f$-value}
&\colhead{$k_{pr}$/10$^{11}$}\\
\colhead{ }
&\colhead{(m\AA)}
&\colhead{(10$^{-3}$)}
&\colhead{(s$^{-1}$)}
&\colhead{ }
&\colhead{(m\AA)}
&\colhead{(10$^{-3}$)}
&\colhead{(s$^{-1}$)}
&\colhead{ }
&\colhead{(m\AA)}
&\colhead{(10$^{-3}$)}
&\colhead{(s$^{-1}$)}}
\startdata
 &\multicolumn{3}{c}{$\lambda$1368.74} & &\multicolumn{3}{c}{$\lambda$1369.13} & &\multicolumn{3}{c}{$\lambda$1370.87}\\
\cline{2-4} \cline{6-8} \cline{10-12}\\
HD 23180 &3.1 $\pm$ 0.2 &20 $\pm$ 1 &1.1 & &5.7 $\pm$ 0.5 &39 $\pm$ 4 &4.3 & &1.1 $\pm$ 0.2 &7 $\pm$ 1 &0.10\tablenotemark{a} \\
HD 23478 &2.9 $\pm$ 0.7 &20 $\pm$ 5 &1.6 & &3.9 $\pm$ 0.7 &26 $\pm$ 5 &2.3 & &1.5 $\pm$ 0.7 &10 $\pm$ 5 &0.09 \\
HD 24398 &2.3 $\pm$ 0.2 &15 $\pm$ 1 &1.0\tablenotemark{a} & &4.4 $\pm$ 0.4 &27 $\pm$ 2 &3.9 & &1.6 $\pm$ 0.2 &9 $\pm$ 1 &0.06 \\
HD 24534 &4.7 $\pm$ 0.1 &16 $\pm$ 1 &0.8 & &8.0 $\pm$ 0.3 &29 $\pm$ 2 &3.8 & &2.2 $\pm$ 0.2 &8 $\pm$ 1 &0.17 \\
HD 147683 &4.5 $\pm$ 0.8 &25 $\pm$ 5 &0.8 & &6.5 $\pm$ 1.7 &38 $\pm$ 11 &4.7 & &1.6 $\pm$ 0.4 &9 $\pm$ 3 &0.10\tablenotemark{a} \\
HD 149757\tablenotemark{b} &5.6 $\pm$ 0.2 &29 $\pm$ 3 &2.1 & &7.5 $\pm$ 0.2 &39 $\pm$ 4 &3.9 & &2.2 $\pm$ 0.3 &11 $\pm$ 2 &0.10\tablenotemark{a} \\
HD 154368\tablenotemark{b} &11.3 $\pm$ 1.2 &24 $\pm$ 3 &1.7 & &17.2 $\pm$ 2.0 &37 $\pm$ 5 &4.1 & &\nodata &\nodata     &\nodata \\
HD 203374 &3.4 $\pm$ 1.2 &19 $\pm$ 7 &1.5 & &5.2 $\pm$ 1.8 &30 $\pm$ 10 &3.3 & &\nodata &\nodata     &\nodata \\
HD 207308 &4.7 $\pm$ 1.0 &19 $\pm$ 4 &2.3 & &5.7 $\pm$ 1.0 &23 $\pm$ 4 &3.8 & &2.5 $\pm$ 0.4 &10 $\pm$ 2 &0.10\tablenotemark{a} \\
HD 207538 &5.5 $\pm$ 0.7 &18 $\pm$ 2 &2.4 & &9.0 $\pm$ 1.3 &31 $\pm$ 5 &4.0 & &2.9 $\pm$ 1.4 &10 $\pm$ 5 &0.10\tablenotemark{a} \\
HD 208266 &4.8 $\pm$ 0.7 &18 $\pm$ 3 &1.7 & &6.9 $\pm$ 1.2 &27 $\pm$ 5 &3.9 & &2.4 $\pm$ 0.5 &9 $\pm$ 2 &0.10\tablenotemark{a} \\
HD 210839\tablenotemark{b} &5.4 $\pm$ 0.5 &33 $\pm$ 4 &1.0\tablenotemark{a} & &7.5 $\pm$ 0.5 &47 $\pm$ 6 &3.6 & &1.7 $\pm$ 0.5 &10 $\pm$ 3 &0.10\tablenotemark{a} \\
\tableline
STIS avg. &        &19 $\pm$ 3 &1.5 $\pm$ 0.6 & &  &30 $\pm$ 5 &3.8 $\pm$ 0.7 & &  &9 $\pm$ 1 &0.11 $\pm$ 0.06 
\enddata
\tablenotetext{a}{These are fixed, not fitted, $k_{pr}$ values.}
\tablenotetext{b}{These GHRS data are not included in the STIS average.}
\end{deluxetable}

\begin{deluxetable}{lccccccccccc}
\tabletypesize{\scriptsize}
\tablewidth{0pc}
\tablecaption{Astronomical Measurements of $4d$$-$$X$, $F$$-$$X$, and $D$$-$$X$ lines of CH}
\tablehead{
\colhead{Star}
&\colhead{W$_{\lambda}$}
&\colhead{$f$-value}
&\colhead{$k_{pr}$/10$^{11}$}
&\colhead{}
&\colhead{W$_{\lambda}$}
&\colhead{$f$-value}
&\colhead{$k_{pr}$/10$^{11}$}
&\colhead{}
&\colhead{W$_{\lambda}$}
&\colhead{$f$-value}
&\colhead{$k_{pr}$/10$^{11}$}\\
\colhead{}
&\colhead{(m\AA)}
&\colhead{(10$^{-3}$)}
&\colhead{(s$^{-1}$)}
&\colhead{}
&\colhead{(m\AA)}
&\colhead{(10$^{-3}$)}
&\colhead{(s$^{-1}$)}
&\colhead{}
&\colhead{(m\AA)}
&\colhead{(10$^{-3}$)}
&\colhead{(s$^{-1}$)}}
\startdata
 &\multicolumn{3}{c}{$4d$$-$$X$ $\lambda$1271.02} & &\multicolumn{3}{c}{$F$$-$$X$ $\lambda$1549.05\tablenotemark{b}} & &\multicolumn{3}{c}{$F$$-$$X$ $\lambda$1549.62} \\
\cline{2-4} \cline{6-8} \cline{10-12}\\
HD 23478 &\nodata&\nodata&\nodata & &5.2 $\pm$ 1.6 &28 $\pm$ 9 &2.0\tablenotemark{a} & &\nodata &\nodata &\nodata \\
HD 24534 &1.6 $\pm$ 0.3 &6 $\pm$ 1 &3.1 & &6.3 $\pm$ 0.9 &17 $\pm$ 2 &3.0 & &3.9 $\pm$ 0.4 &11 $\pm$ 1 &0.6 \\
HD 147683 &\nodata&\nodata&\nodata & &6.6 $\pm$ 1.6 &29 $\pm$ 7 &2.0\tablenotemark{a} & &2.6 $\pm$ 1.5 &15 $\pm$ 9 &2.0\tablenotemark{a} \\
HD 149757 &1.1 $\pm$ 0.1 &6 $\pm$ 1 &3.5 & &4.1 $\pm$ 0.9 &16 $\pm$ 4 &2.5 & &4.4 $\pm$ 0.8 &16 $\pm$ 3 &2.5\tablenotemark{a} \\
HD 203532 &1.3 $\pm$ 0.5 &7 $\pm$ 3 &0.8 & &\nodata&\nodata&\nodata & &\nodata&\nodata&\nodata \\
HD 207308 &2.2 $\pm$ 0.7 &10 $\pm$ 3 &3.0\tablenotemark{a} & &4.6 $\pm$ 0.7 &14 $\pm$ 2 &1.1 & &3.1 $\pm$ 0.9 &9 $\pm$ 3 &1.0\tablenotemark{a} \\
HD 210839 &\nodata&\nodata&\nodata & &4.2 $\pm$ 0.6 &21 $\pm$ 3 &4.3 & &3.0 $\pm$ 0.5 &14 $\pm$ 3 &2.5 \\
\tableline
Average &        &7 $\pm$ 2 &2.5 $\pm$ 1.5 & &  &21 $\pm$ 6 &2.7 $\pm$ 1.3 & & &13 $\pm$ 3 &1.6 $\pm$ 0.9 \\
\tableline\\
 &\multicolumn{3}{c}{$D$$-$$X$ $\lambda$1693.24} & &\multicolumn{3}{c}{$D$$-$$X$ $\lambda$1695.34\tablenotemark{b}} & & & & \\
\cline{2-4} \cline{6-8} \\
HD 24534 &0.6 $\pm$ 0.3 &1.2 $\pm$ 0.6 &0.7\tablenotemark{a} & &1.8 $\pm$ 0.5 &3.8 $\pm$ 1.1 &0.7 & & & & \\
HD 24534\tablenotemark{c} &0.8 $\pm$ 0.4 &1.9 $\pm$ 1.0 &3.6\tablenotemark{a} & &2.5 $\pm$ 1.0 &5.7 $\pm$ 2.3 &3.6\tablenotemark{a} & & & & 
\enddata
\tablenotetext{a}{These are fixed, not fitted, $k_{pr}$ values.}
\tablenotetext{b}{These features are known to be blends of $Q_2$ and $^QR_{12}$, see text.}
\tablenotetext{c}{This $D$$-$$X$ synthesis for HD~24534 incorporates the laboratory value of $\Gamma$ (see text and Figures 5 and 6).}
\end{deluxetable}

\begin{figure}
\epsscale{0.9}
\plotone{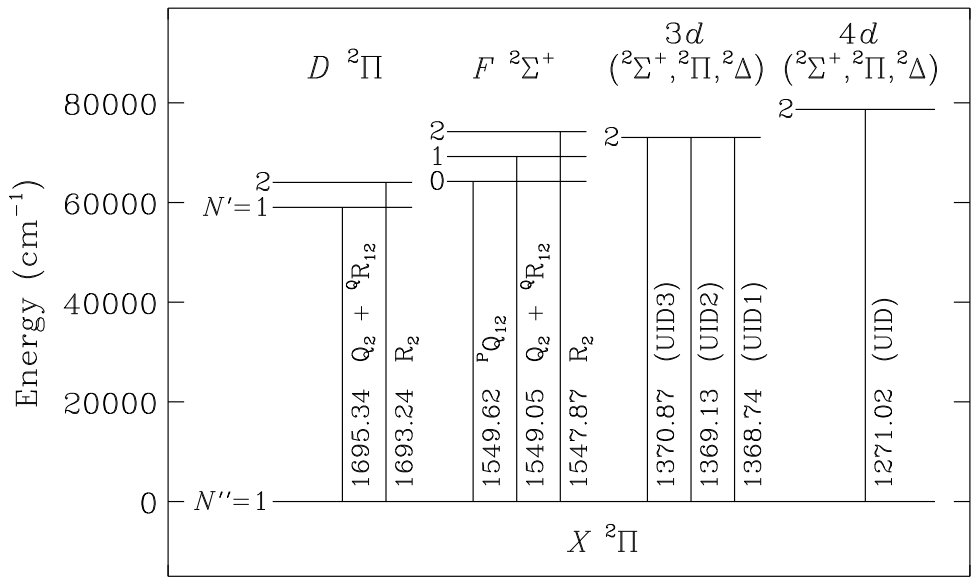}
\caption{Term diagram for electronic states of CH discussed in this paper.
The levels of each state are shown according to their $N$ values, with a schematic not-to-scale
energy separation.
Only the lowest $N^\prime$ level is positioned at the correct location along the energy axis.
This diagram shows only the transitions studied here.
For the 1271.02~\AA\ transition of the 4$d$ Rydberg complex, we assumed that it has the same
$N^\prime$ = 2 that has been assigned by \citet{watson01} to the UID transitions of the 3$d$ Rydberg
complex.}
\end{figure}

\begin{figure}
\epsscale{0.9}
\plotone{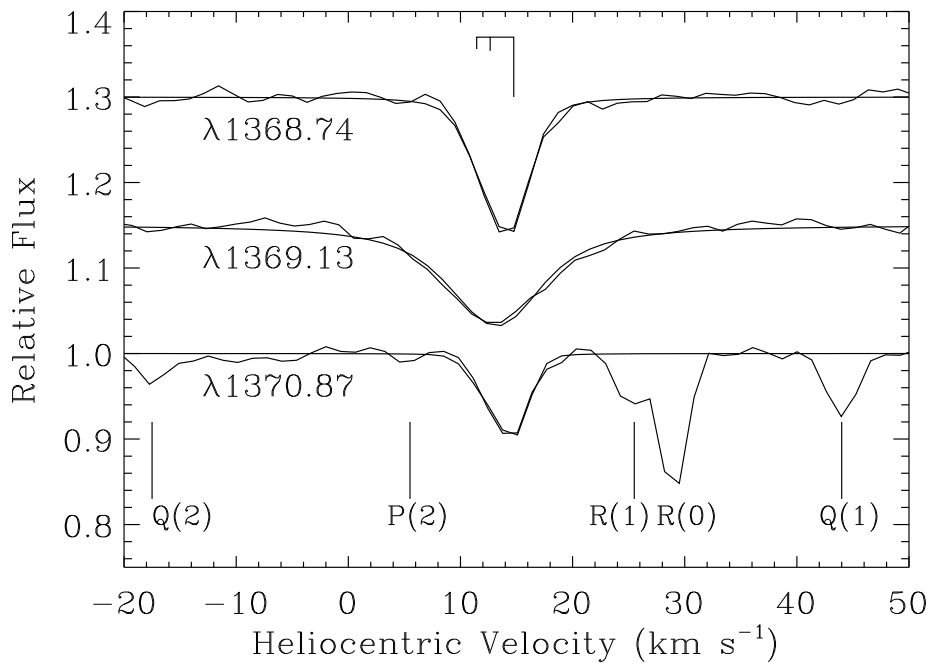}
\caption{Data and fits for $\lambda$$\lambda$1368, 1369, and 1370 from the $3d$$-$$X$ band of CH
toward X~Per, vertically separated by 0.15 continuum units.
This set of data is of higher $R$ and S/N than the spectrum used in the initial detection of these
lines toward $\zeta$~Oph.
In addition, $N$(CH) is higher along the X~Per sight line, which shows a three-cloud component
structure as indicated by the fraction-scaled pointers above the spectra.
The larger width of $\lambda$1369 is obvious.
Its slight blueshift relative to $\lambda$1368 and $\lambda$1370 shows that its rest wavelength
should be adjusted by $\approx$ $-$0.005 \AA\ relative to the sharper lines.
The line $\lambda$1370 is located among lines from $^{13}$CO; those to the blue are from the
$A$$-$$X$ (6$-$0) band, while those to the red are from the $d$$-$$X$ (12$-$0) intersystem band
\protect\citep[see Figure 1 of][]{sfl02}.}
\end{figure}

\begin{figure}
\epsscale{0.9}
\plotone{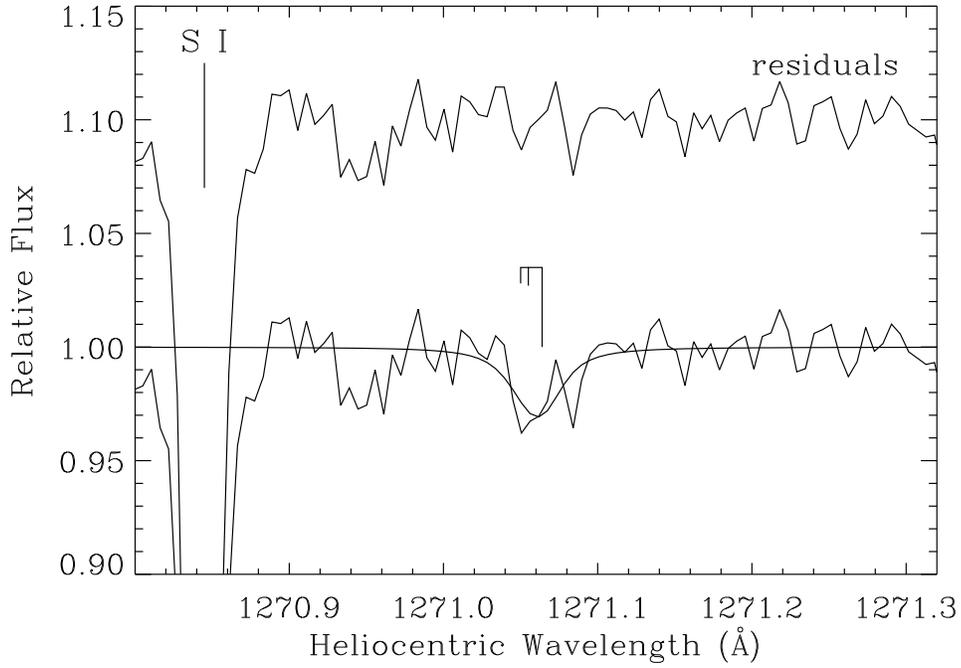}
\caption{Data, fit, and residuals for $\lambda$1271.02 from the $4d$$-$$X$ band of CH
toward X~Per.
The cloud component structure on the line of sight is indicated above the fit.
The strong absorption to the blue is from a triplet of \ion{S}{1} lines at 1270.78~\AA.}
\end{figure}

\begin{figure}
\epsscale{0.9}
\plotone{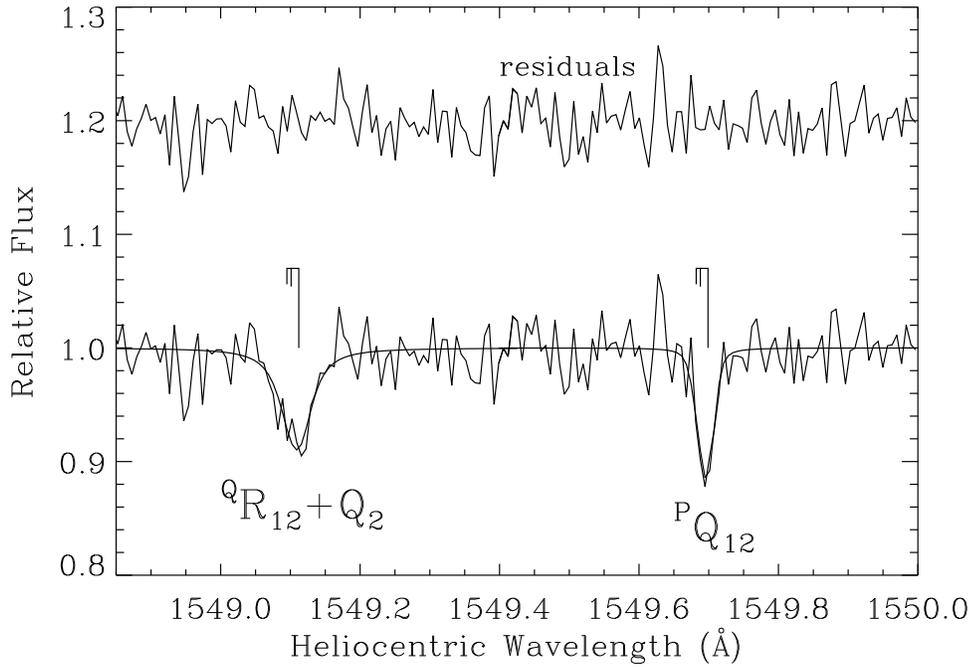}
\caption{Data, fits, and residuals for $\lambda$1549.05 and $\lambda$1549.62 from the $F$$-$$X$
band of CH toward X~Per.
The bluer `line' is a blend of two transitions, $Q_2$ and $^QR_{12}$, but their
separation is unknown.
The redder line is the unblended $^PQ_{12}$ transition.
The cloud component structure is indicated above the features.}
\end{figure}

\begin{figure}
\epsscale{0.9}
\plotone{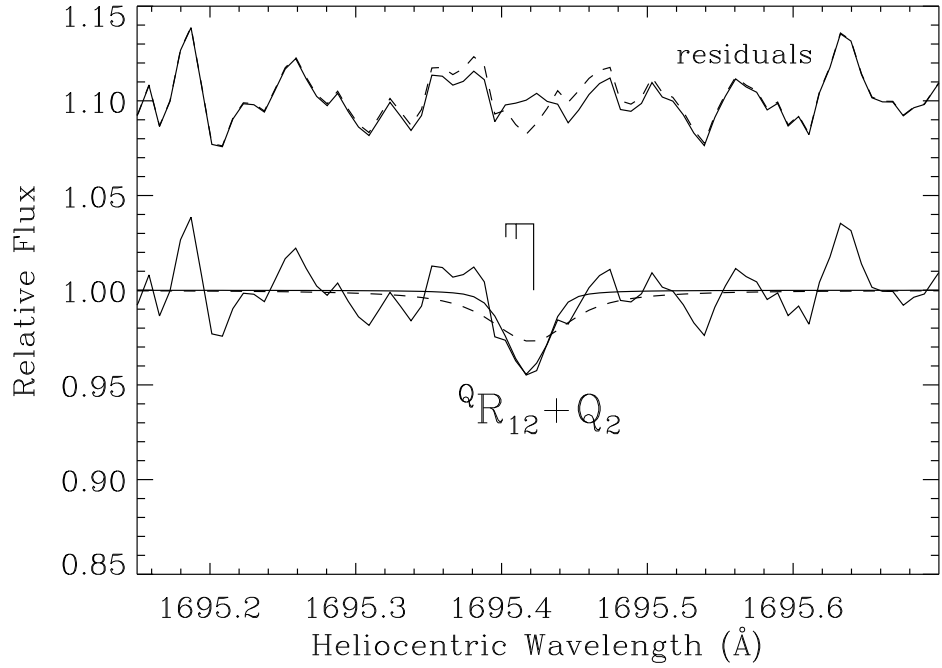}
\caption{Data, fits, and residuals for $\lambda$1695.34 from the $D$$-$$X$
band of CH toward X~Per.
This is a blend of the two transitions $Q_2$ and $^QR_{12}$, the separation
of which is unknown.
The cloud component structure is indicated above the fitted feature.
Our initial fit returned a smaller line width (solid line) than the value measured in the lab
(dashed line), yet both are consistent with the ambient noise in the spectrum.}
\end{figure}

\begin{figure}
\epsscale{0.9}
\plotone{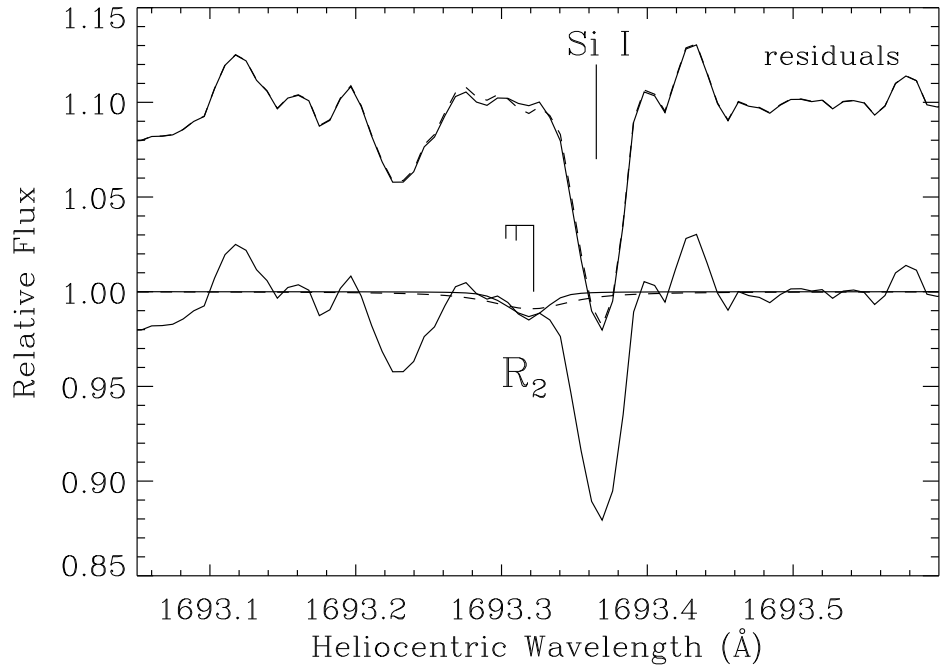}
\caption{Data, fits, and residuals for $\lambda$1693.24 from the $D$$-$$X$
band of CH toward X~Per.
This is the weaker transition $R_2$, which is partially blended with the \ion{Si}{1}
at 1693.29~\AA.
The cloud component structure is indicated above the fitted line.
Both fits employ fixed $\Gamma$ values from the $\lambda$1695.34 fits in Figure 5.
The feature at $\sim$ 1693.25~\AA\ does not appear in any tabulated interstellar line list.}
\end{figure}

\end{document}